\documentclass{elsart3}

   \newcommand{\aap}{A\&A}
   
   \newcommand{\aj}{AJ}

   \newcommand{\apjl}{ApJ}
   \newcommand{\apjs}{ApJS}

\usepackage{natbib}
\usepackage[xdvi]{graphicx}
\usepackage{textcomp}
\usepackage{dcolumn}
\newcolumntype{.}{D{.}{.}{4}}

\usepackage{amssymb}

\begin{document}

\begin{frontmatter}



\title{Molecular gas in QSO host galaxies}


\author[1]{T. Bertram},
\author[1]{A. Eckart},
\author[2]{M. Krips},
\author[1]{C. Straubmeier},
\author[1]{S. Fischer},
\author[3]{J.G. Staguhn}
\address[1]{I. Physikalisches Institut, Universit\"at zu K\"oln, Z\"ulpicher Str. 77, 50937 K\"oln, Germany}
\address[2]{ Smithsonian Astrophysical Observatory (SAO), 
Submillimeter Array (SMA), 645, North A'Ohoku Place, Hilo, HI, 96720, USA}
\address[3]{NASA/Goddard Space Flight Center, Greenbelt, MD 20771, USA}

\begin{abstract}
We present the results of a survey for CO line emission
from a sample of nearby QSO hosts taken from the 
Hamburg/ESO survey (HES) and the V\'eron-Cetty and V\'eron quasar
catalogue. From a total of 39 observed sources we clearly detected 5 objects
with $>$10$\sigma$ signals (HE~0108-4743, HE~0224-2834, J035818.7-612407, HE~1029-1831, HE~2211-3903). Further 6 sources show marginal detections on the 2$\sigma$ level.
\end{abstract}


\end{frontmatter}

\section{Introduction} 
Studies of molecular gas in the host galaxies of QSOs and high rank Seyfert 1 galaxies are essential for the understanding of the star formation in and fueling of the central engines. 
\
Especially the nearby QSOs (z$\le$0.1) represent an important link between the cosmologically local AGN and the high redshift, high luminosity QSOs (z$\ge$0.5). 
\
Observations of galaxies hosting AGN remain challenging, even with state-of-the-art instrumentation. 
\
Studies of this kind require exceptionally high spatial resolutions and sensitivities to be able to separate the nuclear component from the faint contribution of the underlying galaxy. 
\
But even with the highest resolutions feasible, a detailed analysis of the distribution and kinematics of the molecular gas component is possible only for the closest objects. 
\
There are only few cases which have been investigated so far, among them 3C48  \citep[for details cf.][and references therein]{zuther2005}.
\
For studies of further objects the identification of suitable sources is mandatory.
\
Until today, only a few surveys have been carried out, which are dedicated to the detection of molecular gas in QSOs   \citep[e.g.][]{2001aj....121.3285e,evans2005,scoville2003}.

\section{The Sample}
The sources in our sample were selected from the Hamburg/ESO survey (HES) \citep{2000a&a...358...77w} for bright QSOs and the V\'eron-Cetty and V\'eron quasar catalogue \citep{2001a&a...374...92v}. 
\
Unlike preceding surveys, HES does not discriminate against extended sources, thus enabling the study of host galaxies.
\
The only selection criterion for the nearby QSO sample was the cosmological distance: only objects with a redshift z$<$0.060 were chosen. 
\
This redshift limit ensures the observability of the important diagnostic  CO(2-0) rotation vibrational band head absorption line, which is then still accessible in the K-band \citep[cf.][]{fischer2005}. 
\
The sample consists of 63 sources.
\
It is entirely based on a volume limit and not based on a FIR flux selection criterion.
\section{Results}
We scanned 39 members of the nearby QSO sample for millimetric CO emission so far. 
\
The primary goal was to identify the CO brightest objects for high resolution interferometric followup observations. 
\
These initial measurements were carried out in part with the BIMA array, in part with the SEST 15m single-dish telescope and resulted in 5 detections.
\
Another 6 sources show marginal detections on the 2s level and deserve closer investigation in the near future. 
\
Table \ref{TabSESTResults} lists the integrated temperatures, derived CO luminosities and molecular gas mass estimates for all objects observed with SEST.
\
The CO luminosities were determined, using
$$ {\rm L^\prime_{CO}=23.5 \;\Omega_{B}\;D_L^2 \; I_{CO}\; (1+z)^{-3}\;(K\; km\; s^{-1}\; pc^2),}$$ \citep{1992apj...398l..29s}, where ${\rm \Omega_{B}}$ is the telescope beam (45'').
\
The luminosity distance D$_{\rm L}$ was calculated assuming 
H$_0$=75km s$^{-1}$ Mpc$^{-1}$ and q$_0$=0.5. 
\
All upper limits are based on 3$\sigma$  antenna temperature limits and a mean linewidth of  280~km~s$^{-1}$.
\
To derive molecular gas mass estimates, a conversion factor $\alpha=4$M$_{\odot}($K km s$^{-1}$ pc$^2$)$^{-1}$ was applied \citep[][and references therein]{2001aj....121.3285e,scoville2003}.
\\

In the case of HE1029-1831 the BIMA survey led to a detection of bright CO line emission in a compact region. 
\
The medium spatial resolution of 13.6'' $\times$ 5.8'' that was achieved in this antenna array configuration did not allow to resolve the source. 
\
This was achieved by follow-up PdBI observations in B and C configuration in Feb. 2002 and in A configuration in Mar. 2003.
\
These observations revealed a strong association of the detected CO(1-0) and CO(2-1) line emission with the optical bar. 
\
In the position-velocity diagrams a strong velocity gradient across the bar indicates a bar-driven inflow of gas. 
\
The CO emission is consistent with a simple bar model. 
\
The CO(2-1)/CO(1-0) line ratio is estimated to be ~0.7, which indicates subthermally excited, cold gas typically found in a disk. 
\
A detailed analysis will be presented in \citet{krips}.
\\
\begin{figure*}
  \centering
    \resizebox{\hsize}{!}{\includegraphics{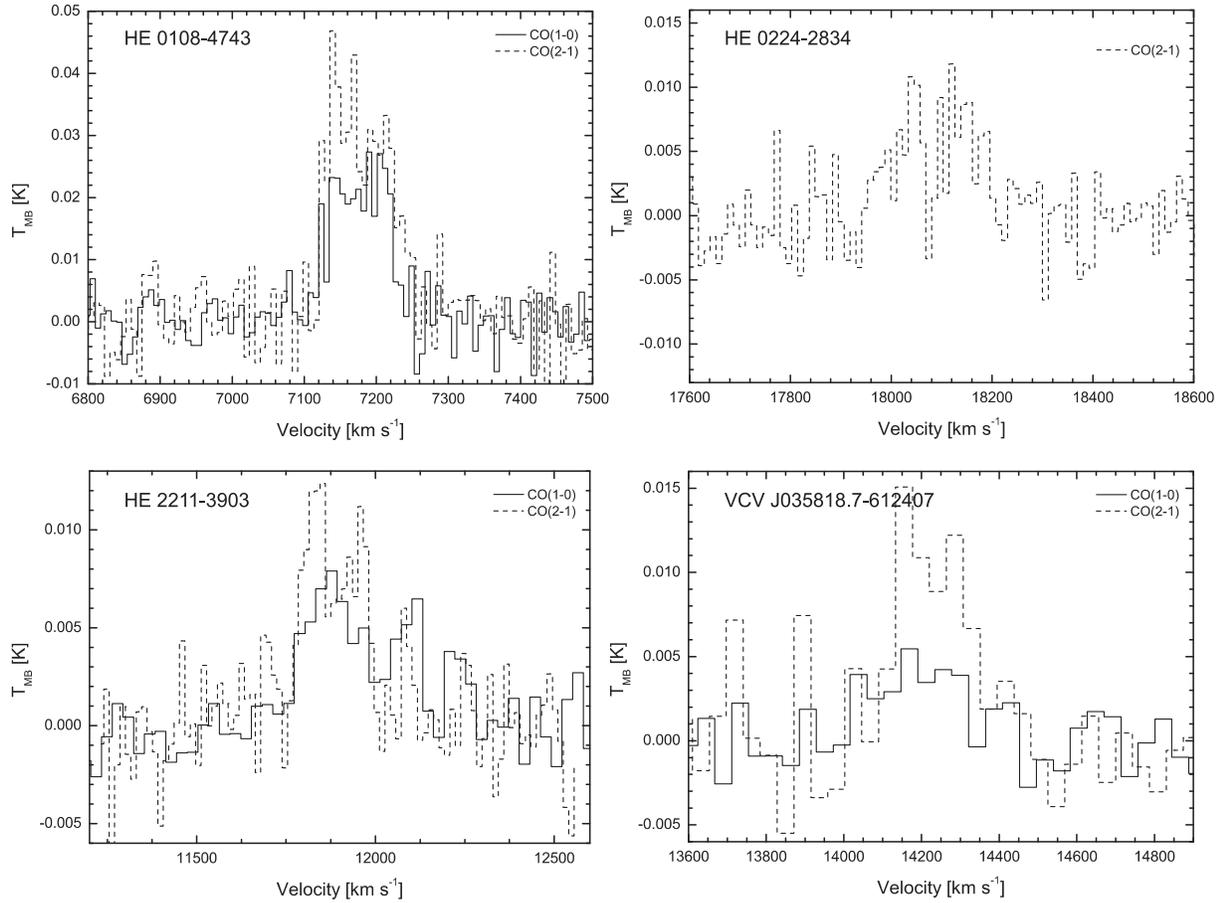}}
  \caption[]{$^{12}$CO spectra of 4 members of the QSO sample detected with SEST. }
  \label{FigSESTSpectra}
\end{figure*}

\begin{table*}
  \caption[]{Measured $^{12}$CO properties of members of the QSO sample 
             obtained with SEST.
	     The Objects marked with "*" show weak signals which need to be confirmed.
	     For the upper limits a linewidth of 280 km s$^{-1}$ was assumed.
	     I$_{\rm CO}$ was determined for re-sampled spectra with resolutions given below.}     
  \begin{tabular}{lr p{0.3cm} .. p{0.3cm}..p{0.3cm}.p{0.3cm}. }
     \hline
     \hline
     \multicolumn{1}{c}{Obj.}              & 
     \multicolumn{1}{c}{cz}                &
                                           &
     \multicolumn{1}{c}{spectral Res.}     &
     \multicolumn{1}{c}{I$_{\rm CO(1-0)}$ }& 
                                           &
     \multicolumn{1}{c}{spectral Res.}     &  
     \multicolumn{1}{c}{I$_{\rm CO(2-1)}$} &
                                           &
     \multicolumn{1}{c}{L$^\prime_{\rm CO}$} &
                                           &
     \multicolumn{1}{c}{M(H$_2)$}         \\
                                           & 
     \multicolumn{1}{c}{[km s$^{-1}$]}	   &
                                           &
     \multicolumn{1}{c}{[km s$^{-1}$]}     &
     \multicolumn{1}{c}{[K\, km s$^{-1}$]} & 
                                           &
     \multicolumn{1}{c}{[km s$^{-1}$]}     &
     \multicolumn{1}{c}{[K\, km s$^{-1}$]} &
                                           &
     \multicolumn{1}{c}{[10$^9 $K km s$^{-1}$ pc$^2$]} &
                                           &
     \multicolumn{1}{c}{[10$^9$ M$_{\odot}$]}\\ 
    \hline
    HE 0003-5023*             & 10027& & 14.9 & < 0.8       & & 15.0 & < 0.8       & &    < 0.71   & &<  2.83   \\
    HE 0036-5133              & 8640 & & 29.6 & < 1.2       & & 15.0 & < 0.8       & &    < 0.71   & &<  2.83   \\
    HE 0051-2420              & 16788& & 22.0 & < 1.0       & & 22.0 & < 1.0       & &    < 1.88   & &<  7.54   \\
    J005924.5+270332          & 13770& & 21.8 & < 1.0       & & 21.7 & < 1.9       & &    < 1.31   & &<  5.25   \\
    HE 0103-3447              & 17088& & 30.5 & < 1.2       & & 7.7  & < 0.6       & &    < 2.59   & &< 10.36   \\
    HE 0108-4743              & 7285 & &  7.4 & 2.2\pm 0.1  & & 7.5  & 3.8 \pm 0.1 & &0.97 \pm 0.04& &   3.87   \\
    J012345.8\textminus584821 & 14095& & 15.1 & < 0.6       & & 15.2 & < 0.8       & &    < 0.92   & &<  3.68   \\
    HE 0122-5137              & 15589& & 30.3 & < 0.8       & & 15.3 & < 0.8       & &    < 1.64   & &<  6.55   \\
    HE 0224-2834              & 17940& & 15.3 & < 0.3       & & 11.6 & 1.3 \pm 0.1 & &0.71 \pm 0.05& &   2.85   \\
    HE 0323-4204              & 17388& & 30.5 & < 0.8       & & 15.4 & < 0.8       & &    < 2.06   & &<  8.25   \\
    HE 0336-5545*             & 17688& & 30.5 & < 0.4       & & 30.8 & < 1.1       & &    < 0.70   & &<  2.78   \\
    HE 0343-3943              & 12933& & 30.0 & < 1.2       & & 15.2 & < 0.8       & &    < 1.54   & &<  6.16   \\
    HE 0349-4036*             & 17440& & 15.2 & < 0.8       & & 15.4 & < 0.4       & &    < 2.02   & &<  8.09   \\
    J035818.7\textminus612407 & 14264& & 43.6 & 1.3\pm 0.1  & & 43.6 & 3.0 \pm 0.1 & &2.02 \pm 0.15& &   8.07   \\
    HE 0359-3841              & 17688& & 22.0 & < 1.0       & & 22.0 & < 1.5       & &    < 2.08   & &<   8.3   \\
    HE 0403-3719              & 16540& & 15.2 & < 0.6       & & 15.4 & < 0.4       & &    < 1.22   & &<  4.89   \\
    HE 0429-5343              & 11994& & 30.0 & < 0.4       & & 30.3 & < 0.6       & &    < 0.33   & &<  1.33   \\
    HE 0436-4717*             & 15889& & 30.3 & < 0.8       & & 15.3 & < 0.4       & &    < 1.69   & &<  6.78   \\
    HE 0535-4224              & 10493& & 14.9 & < 0.6       & & 15.1 & < 0.8       & &    < 0.52   & &<  2.07   \\
    J061320.8\textminus324154 & 14990& & 21.9 & < 1.0       & & 21.9 & <1.0        & &    < 1.58   & &<  6.33   \\
    HE 0853-0126              & 17930& & 22.1 & < 1.0       & & 22.1 & < 1.5       & &    < 2.14   & &<  8.56   \\
    J091609.5\textminus621929 & 17178& & 22.0 & < 0.7       & & 22.0 & < 1.5       & &    < 1.30   & &<  5.21   \\
    HE 1013-1947              & 16413& & 21.9 & < 1.0       & & 21.9 & < 1.9       & &    < 1.82   & &<  7.27   \\
    J204409.7\textminus104324 & 10312& & 10.8 & < 0.9       & & 21.5 & < 0.9       & &    < 0.75   & &<  2.99   \\
    HE 2211-3903              & 11906& & 30.0 & 1.9 \pm 0.1 & & 15.1 & 1.9 \pm 0.1 & & 2.1 \pm 0.11& &   8.39   \\
    HE 2231-3722              & 12891& & 15.0 & < 0.8       & & 15.2 & < 0.8       & &    < 1.14   & &<  4.56   \\
    HE 2236-3621              & 17988& & 22.1 & < 1.0       & & 22.1 & < 1.5       & &    < 2.13   & &<  8.54   \\
    J230443.5\textminus084108 & 14185& & 21.8 & < 1.0       & & 21.8 & < 1.9       & &    < 1.39   & &<  5.57   \\
    HE 2354-3044*             & 9198 & & 29.7 & < 1.2       & & 30.0 & < 1.1       & &    < 0.80   & &<   3.2   \\
                                                                                    
    \hline                   
    \end{tabular} 
  \label{TabSESTResults}
\end{table*}

In Sept. 2005 we were able to observe further 15 objects from an extended HES list with the IRAM 30 m telescope simultaneously in CO(1-0) and CO(2-1). With the extended list, it is possible to restrict upcoming analyses of the molecular gas content and distribution of AGN hosts with z$<$0.06 to HES objects and hence gain in homogeneity. With the greater sensitivity of the IRAM 30m telescope, the goal was to observe to a common L$^\prime_{\rm CO}$ limit of 2.5$\cdot$10$^{8}{\rm K\; km\; s^{-1}\; pc^2}$. These observations resulted in a total of 10 detections. A detailed analysis will follow in an upcoming paper \citep{bertram}.        

\section{Conclusion}
To get an unbiased result, no preselection based on IRAS flux densities was applied. Out of 13 observed IRAS sources we detected 5 in their CO mm-line emission resulting in a ~40\% detection rate. 2 of the 6 marginally detected sources are IRAS sources, too. Including them, the detection rate amongst the IRAS sources is approximately 55\%. This is the same order of magnitude as found by Evans et al. (2001) and also comparable to the detection rate reported by Scoville et al. (2003). Out of a total of 39 observed sources the 5 detected IRAS sources represent $\sim$ 13\%. This result indicates that not all QSOs/high rank Seyferts reside in gas rich hosts as suggested by Scoville et al. (2003). 
Only one of the detected QSO hosts is identified as an elliptical, four of them are identified as spirals, two of them as barred spirals.  Among the remaining non-detections the percentage of ellipticals (at least 26 sources) and S0 galaxies (at least 10 sources) is high. Only 12 are clearly identified as spirals, 9 of which are barred spirals. Although the 3$\sigma$ upper limits for the molecular gas masses are high, this source classification is consistent with their preferential association with elliptical galaxies.

\label{}




\end{document}